\documentclass[preprint,preprintnumbers,nofootinbib,aps,superscriptaddress,10pt,twocolumn]{revtex4-1}       
\usepackage{amsmath,amssymb,bm,epsfig}
\usepackage{color}
\usepackage{natbib}
\usepackage{hyperref}
\usepackage{ulem}
\usepackage{graphicx}
\usepackage{xcolor,colortbl}
\usepackage{pifont}
\usepackage{float}
\usepackage{dcolumn}
\usepackage{booktabs}
\usepackage{multirow}

\long\def\/*#1*/{}

\def \beq{\begin{equation}}
\def \eeq{\end{equation}}
\def \beqa{\begin{eqnarray}}
\def \eeqa{\end{eqnarray}}
\usepackage{xspace}

\newcommand{\sNN}{\sqrt{s_{\rm NN}}}

\newcommand{\pt}{p_{\rm T}}

\def \beq{\begin{equation}}
\def \eeq{\end{equation}}
\def \beqa{\begin{eqnarray}}
\def \eeqa{\end{eqnarray}}

\begin{document}

\title{Multiplicity dependence freeze-out scenarios in pp collisions at $\sqrt{s}$ = 7 TeV}

\author{Susil Kumar Panda}
 \email{susilkpanda1@gmail.com}
 \affiliation{Department of Physics,Ravenshaw University, Cuttack - 753003, India}

\author{Sandeep Chatterjee}
 \email{Sandeep.Chatterjee@fis.agh.edu.pl}
 \affiliation{Department of Physical Sciences, Indian Institute of Science Education and Research, Berhampur, Transit Campus, 
Government ITI, Berhampur 760010, Odisha, India}
 
 \author{Ajay Kumar Dash}
 \email{ajayd@niser.ac.in}
 \affiliation{School of Earth and Planetary Sciences, National Institute of Science 
Education and Research, HBNI, Jatni - 752050, India}
 
 \author{Bedangadas Mohanty}
 \email{bedanga@niser.ac.in}
 \affiliation{School of Physical Sciences, National Institute 
 of Science Education and Research, HBNI, 
 Jatni, 752050, India}
 
 \author{Rita Paikaray}
 \email{r\_paikaray@rediffmail.com}
 \affiliation{Department of Physics,Ravenshaw University, Cuttack - 753003, India} 
 
 \author{Subhasis Samanta}
 \email{subhasis.samant@gmail.com}
 \affiliation{Institute of Physics, Jan Kochanowski University, 25-406 Kielce, Poland} 
 
 \author{Ranbir Singh}
 \email{ranbir.singh@niser.ac.in}
 \affiliation{School of Physical Sciences, National Institute 
 of Science Education and Research, HBNI, 
 Jatni, 752050, India}

\begin{abstract}

The data on transverse momentum integrated hadron yields in different multiplicity classes of p+p collisions 
at $\sqrt{s}=7$ TeV have been analyzed to extract the chemical freeze-out parameters using a thermal model.
The chemical freeze-out parameters have been extracted for three different freeze-out schemes:
i. unified freeze-out for all hadrons in complete thermal equilibrium (1CFO), 
ii. unified freeze-out for all hadrons with an additional parameter $\gamma_S$ which accounts for possible 
out-of-equilibrium production of strange hadrons (1CFO$+\gamma_S$), and 
iii. separate freeze-out for hadrons with and without strangeness content (2CFO).
It has been observed that 1CFO$+\gamma_S$ scheme gives the best description of the hadronic yields at midrapidity
when multiplicity ($\langle dN_{ch}/d\eta \rangle$) of the collision is less than 10. 
This indicates that the strangeness is out of equilibrium in most of the multiplicity classes of p+p collisions.
All the three parameters of this CFO scheme, temperature ($T$), radius of the fireball ($R$) and 
strangeness suppression factor ($\gamma_S$) increase with the increase of $\langle dN_{ch}/d\eta \rangle$.
Further, we have compared applicability of different CFO schemes considering two more colliding system
p+Pb at $\sNN$ = 5.02 and Pb+Pb at $\sNN$ = 2.76 TeV along with p+p collisions at $\sqrt{s}=7$ TeV.
We observe a freeze-out volume (or multiplicity) dependence of CFO schemes regardless of colliding ions.
The 1CFO+$\gamma_S$, 1CFO and 2CFO schemes provide
best description of the data when the dimension less quantity $VT^3$ approximately
satisfies the conditions $VT^3 <50$, $50 < VT^3 < 100$ and $VT^3 > 100$ respectively or the corresponding
multiplicity satisfies the conditions $\langle dN_{ch}/d\eta \rangle<30$,  $30 < dN_{ch}/d\eta < 60$ and $\langle dN_{ch}/d\eta \rangle>100$ respectively.


\end{abstract}
\maketitle

\section{Introduction}\label{sec.intro}

In a high-energy ion collision, a fireball is produced.
When the energy density at the core of the fireball is 
sufficiently high the quark-gluon plasma (QGP), a deconfined phase of quarks and gluons
is formed. The fireball expands resulting in a decrease in the temperature.
A QGP to hadronic phase transition occurs when the temperature
drops below the transition temperature. At the initial stage, hadrons interact
both elastically and inelastically in the hadronic medium. 
Inelastic interaction among the hadrons ceases
at chemical freeze-out (CFO). At this stage, hadronic yields get fixed and do not
change afterwards.
The elastic interaction continues until the kinetic freeze-out is reached.
Hadrons then stream outwards freely and eventually reach the detector.
Experimental data on hadronic yields in ion collisions are traditionally described by
the hadron resonance gas (HRG) model.
In the most simplified formulation of the HRG model, it is assumed
that the CFO of all the hadrons and resonances occurs at the same temperature, baryon chemical potential
and volume of the freeze-out surface. We call this unified CFO scheme 1CFO. 
This 1CFO HRG model successfully describes the hadron yields in nucleus-nucleus collision
across a wide range of centre-of-mass energy
\cite{BraunMunzinger:1995bp,Yen:1998pa,BraunMunzinger:1999qy, Andronic:2005yp}.
To describe strange hadrons in elementary collisions like
$e^+ e^-, pp$ and $p \bar{p}$ \cite{Becattini:1995if,Becattini:1995xt,Becattini:1995if,Becattini:1997rv},
an additional parameter $\gamma_s$ was introduced. 
The parameter $\gamma_s$ accounts for the deviation from chemical equilibrium in the strange sector.
This CFO scheme is referred to as 1CFO+$\gamma_s$.
The Pb-Pb collision data at 2.76 TeV indicated a separation of CFO
between light and strange quark hadrons \cite{Abelev:2012wca}.
The lattice QCD study also suggested a separate CFO for the strange hadrons \cite{Bellwied:2013cta}.
Later, a flavor dependent sequential freeze-out scheme with different freeze-out 
surfaces for hadrons with zero and non-zero
strangeness content was proposed \cite{Chatterjee:2013yga}. This CFO scheme is referred to as 2CFO.
Thus, despite the phenomenological success of HRG models, 
it is not clear which CFO scheme is suitable for which data.
As a result, the understanding of thermal and chemical equilibration is still 
an open issue.
In Ref. \cite{Chatterjee:2016cog} it was shown that hadron yield for 
Pb+Pb collision at $\sNN$ = 2.76 TeV \cite{Abelev:2013vea,Abelev:2013xaa,ABELEV:2013zaa,Abelev:2014uua} 
can be described well with the 2CFO scheme
whereas minimum bias events of p+p collision at $\sqrt{s}$ = 7 TeV \cite{Abelev:2012hy,Abelev:2012jp,Adam:2015qaa} prefer
1CFO+$\gamma_S$ scheme. 
In this present work, we have done a similar
analysis using the newly available data of multiplicity dependence of hadron yield  
created in p+p collision at $\sqrt{s}$ = 7 TeV \cite{Acharya:2018orn}.
We have also compared multiplicity dependence of different CFO schemes at LHC considering p+p collisions at 
$\sqrt{s}=7$ TeV, p+Pb collisions at $\sNN$ = 5.02 and Pb+Pb collisions at $\sNN$ = 2.76 TeV.
Some recent work on multiplicity dependence of freeze-out parameters at the LHC energy
can be found in Refs. \cite{Sharma:2018jqf,Vovchenko:2019kes}.

The paper is arranged in the following way. In Sec.~ \ref{mod} we discussed 
different CFO schemes used for this study. 
The results from the model and data are compared in Sec.~\ref{res}. Finally, we summarise our findings in 
Sec.~\ref{sum}.

 \section{Different chemical freeze-out schemes}\label{mod}
\subsection{1CFO scheme}
In the 1CFO scheme, it is assumed that the fireball at the chemical freeze-out is in thermal and chemical equilibrium.
The logarithm of the partition function of multi-component gas of hadrons and resonances in the  grand canonical ensemble can be
written as
\begin {equation}
 \ln Z=\sum_i \ln Z_i,
\end{equation}
where the sum is over all the species.
For  $i$-th hadrons
\begin{equation}
 \ln Z_i=\pm \frac{V g_i}{2\pi^2}\int_0^\infty p^2\,dp \ln[1\pm\exp(-(E_i-\mu_i)/T)],
\end{equation}
where the upper and lower sign of $\pm$ corresponds to fermions and bosons, respectively.
In the last expression, $V$ is the fireball volume, $T$ is the chemical freeze-out
temperature. 
$g_i$, $m_i$, $\mu_i$ are the degeneracy factor, mass, 
chemical potential of the $i$-th hadron respectively.
The chemical potential of the $i$-th hadron
can be written as $\mu_i = B_i \mu_B + S_i \mu_S + Q_i \mu_Q$ where
$B_i,S_i, Q_i$ are respectively the baryon number, strangeness and 
electric charge of the hadron.
All the chemical potentials are not independent.
The $\mu_S$ and $\mu_Q$ can be extracted applying the following constraints
\beqa
\text{Net} S &=& 0\label{eq.nets}\\
\text{Net} B/\text{Net} Q &=& r \label{eq.netq}
\eeqa
For A+A collisions, the value of $r$ is $\sim2.5$ while it is 1 for p+p collisions.
These constituents come from the conservation of baryon, strangeness and charge quantum numbers of the colliding ions.

The yield of the $i$-th hadron in the grand canonical ensemble can be written as
\begin{equation}\label{eq:n}
N_i = T \frac{\partial \ln Z_i}{\partial \mu_i} = \frac{g_i V}{2 \pi^2} \sum_{k = 1}^{\infty} (\pm)^{k+1} \frac{m_i^2 T}{k} K_2\left(\frac{km_i}{T}\right)e^{\frac{k\mu_i}{T}},
\end{equation}
where $K_2$ is the modified Bessel function of the second kind.
In the last expression, the plus sign is for bosons and the minus sign is for fermions.

Hadronic yields measured by the detectors in experiments
include feed-down from heavier hadrons and resonances. 
Therefore, the total hadronic yield is obtained by including the resonance decay 
contribution to the above primordial
yield
\begin{equation}\label{Eq:yield1CFO}
 N_i^{total} = N_i + \sum_j B.R._{ij} N_j,
\end{equation}
where $B.R._{ij}$ is the branching ratio for the $j$-th hadron species
to $i$-th hadron species.

\subsection{1CFO + $\gamma_S$ scheme}
In this CFO scheme, incomplete strangeness equilibration is assumed.
The hadronic yield in this CFO scheme can be written as \cite{Das:2016muc}
\begin{equation}\label{Eq:yield1CFO_gammaS}
N_i = \frac{g_i V}{2 \pi^2} \sum_{k = 1}^{\infty} (\pm)^{k+1} \frac{m_i^2 T}{k} K_2\left(\frac{km_i}{T}\right)e^{\frac{k\mu_i}{T}} \gamma_S^{k |S_i|},
\end{equation}
where $|S_{i}|$ is the number of valence strange quarks  and
anti-quarks in the hadron species $i$. 
The value $\gamma_S = 1$ corresponds to complete
strangeness equilibration. Compared to the 1CFO scheme (Eq. \ref{eq:n}),
only $\gamma_S$ factor is extra here. 

\subsection{2CFO scheme}
The expression of hadronic yield in this CFO scheme is as Eq. \ref{Eq:yield1CFO}.
However, there are two different sets of parameter for 
hadrons with and without strangeness content.

Within the GCE ensemble, conservation laws for quantum or particle numbers are 
enforced on average through the temperature and chemical potentials. It allows fluctuations of conserved averages.

\subsection{Canonical and strangeness canonical ensemble}
Apart from the grand canonical ensemble, people have also used other ensemble like canonical and strangeness canonical ensemble.
In canonical ensemble (CE) all the charges are exactly conserved (CE)~\cite{Wheaton:2004qb} whereas in strangeness canonical ensemble (SCE)~\cite{Wheaton:2004qb,Kraus:2008fh}
only strangeness is exactly conserved. Both CE and SCE are implemented in the THERMUS program~\cite{Wheaton:2004qb}.

The available experimental data are measured in unit rapidity, so in this work the exact conservation is enforced across only a single unit of rapidity.

\section{Result}\label{res}
\subsection{Multiplicity dependence freeze-out scenarios in p+p collision at LHC}

\begin{figure}[!hbtb]
 \begin{center}
\includegraphics[scale=0.4]{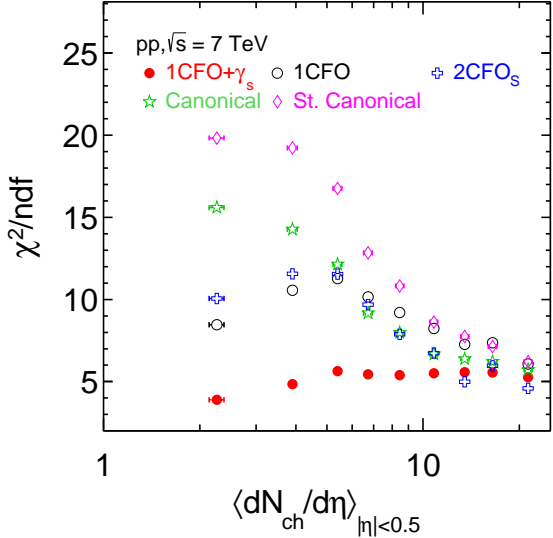}
\caption{(Color online) 
 Variation of $\chi^2$/ndf for the three different freeze-out schemes 
 (1CFO, 1CFO$+\gamma_S$ and 2CFO) with the charge particle multiplicity measured at midrapidity in p+p collisions at 
 $\sqrt{s}$ = 7 TeV \cite{Acharya:2018orn}.
 }
 \label{fig:chisqr_pp}
 \end{center}
\end{figure}

\begin{table*}[]
 \centering
 \caption{The chemical freeze-out parameters in the 1CFO, 1CFO+$\gamma_{s}$ and
 2CFO schemes  in p+p collisions at $\sqrt{s}$ = 7 TeV for for different
 multiplicity classes.}
 \label{tab:fitparameter}
 \begin{tabular}{@{}|c|c|c|c|c|c|c|c|c|@{}}
 \toprule
\hline

 Centrality (\%) & \multicolumn{3} {c|} {1CFO (1CFO+$\gamma_{s}$)}  & \multicolumn{4}{c|}{2CFO} \\ \cmidrule(l){5-8} 
\cline{5-8}
 & \multicolumn{3}{c|}{} & \multicolumn{2}{c|}{Strange}    & \multicolumn{2}{c|}{Non-Strange} \\ \cmidrule(l){2-8}
\cline{2-8} 
                   & $T$  (MeV)     & $R$ (fm)                  & $\gamma_{s}$   & $T_{S}$ (MeV)& $R_{S}$ (fm)& $T_{NS}$  (MeV) & $R_{NS}$ (fm)\\ \hline
0-1             &162$\pm$ 3  (164$\pm$ 3) & 2.29 $\pm$ 0.10 (2.30$\pm$ 0.11) &  0.89 $\pm$ 0.03 &167$\pm$ 3 &2.06$\pm$ 0.10 &145$\pm$ 4 &3.16$\pm$ 0.20  \\ \hline
1-5             &160$\pm$ 3  (163$\pm$ 3) & 2.15 $\pm$ 0.09 (2.16$\pm$ 0.10) &  0.86 $\pm$ 0.03 &165$\pm$ 3 &1.93$\pm$ 0.09 &145$\pm$ 3 &2.90$\pm$ 0.15  \\ \hline
5-10           &161$\pm$ 3  (163$\pm$ 3) & 1.99 $\pm$ 0.09 (2.01$\pm$ 0.09) &  0.86 $\pm$ 0.03 &166$\pm$ 3 &1.78$\pm$ 0.09 &145$\pm$ 3 &2.71$\pm$ 0.13  \\ \hline
10-20         &159$\pm$ 3  (162$\pm$ 3) & 1.89 $\pm$ 0.08 (1.91$\pm$ 0.08) &  0.84 $\pm$ 0.03 &164$\pm$ 3 &1.70$\pm$ 0.09 &146$\pm$ 3 &2.52$\pm$ 0.14  \\ \hline
20-30         &158$\pm$ 3  (161$\pm$ 3) & 1.77 $\pm$ 0.08 (1.80$\pm$ 0.08) &  0.82 $\pm$ 0.03 &163$\pm$ 3 &1.60$\pm$ 0.08 &146$\pm$ 3 &2.34$\pm$ 0.13  \\ \hline
30-40         &157$\pm$ 3  (160$\pm$ 3) & 1.69 $\pm$ 0.07 (1.71$\pm$ 0.08) &  0.80 $\pm$ 0.03 &161$\pm$ 3 &1.52$\pm$ 0.08 &146$\pm$ 3 &2.18$\pm$ 0.12  \\ \hline
40-50         &156$\pm$ 3  (159$\pm$ 3) & 1.58 $\pm$ 0.07 (1.62$\pm$ 0.07) &  0.78 $\pm$ 0.03 &160$\pm$ 3 &1.43$\pm$ 0.07 &145$\pm$ 3 &2.05$\pm$ 0.12  \\ \hline
50-70         &153$\pm$ 3  (157$\pm$ 3) & 1.47 $\pm$ 0.07 (1.51$\pm$ 0.07) &  0.76 $\pm$ 0.03 &157$\pm$ 3 &1.35$\pm$ 0.07 &144$\pm$ 4 &1.88$\pm$ 0.12  \\ \hline
70-100       &148$\pm$ 3  (152$\pm$ 3) & 1.33 $\pm$ 0.07 (1.38$\pm$ 0.07) &  0.72 $\pm$ 0.04 &149$\pm$ 3 &1.27$\pm$ 0.07&140$\pm$ 4  &1.69$\pm$ 0.13  \\ \hline
\end{tabular}                                                                                                                       
\end{table*}

\/*
\begin{table*}[]
\centering
 \caption{The chemical freeze-out parameters in the CE and SCE in p+p collisions at $\sqrt{s}$ = 7 TeV for for different
 multiplicity classes.}
 \label{tab:fitparameterCE}
 \begin{tabular}{@{}|c|c|c|c|c|c|c|c|@{}}
  \toprule
\hline
 Centrality (\%) & \multicolumn{3} {c|} {CE}   & \multicolumn{3} {c|} {SCE} \\
\cline{2-7} 
& $T$  (MeV)     & $R$ (fm)                  & $\gamma_{s}$ & $T$  (MeV)     & $R$ (fm)                  & $\gamma_{s}$ \\
\hline
0-1 & $173 \pm 3$ & $2.06 \pm 0.11$ & $0.91 \pm 0.03$ & $167 \pm 3$ & $2.17 \pm 0.11$ & $0.97 \pm 0.03$\\
\hline
1-5 & $175 \pm 3$ & $1.87 \pm 0.09$ & $0.87 \pm 0.03$ & $167 \pm 3$ & $ 2.01 \pm 0.10$ & $0.96 \pm 0.03$ \\
\hline
5-10 & $178 \pm 3$  & $1.69 \pm 0.09 $ &  $0.89 \pm 0.03$& $168 \pm3$ & $1.85 \pm 0.08$ & $0.98 \pm 0.03$  \\      
\hline

10-20 & $180 \pm 3$ & $1.56 \pm 0.08$ & $0.87 \pm 0.03 $ & $167 \pm 3$ & $1.76 \pm 0.09$ & $0.97 \pm 0.03$    \\      
\hline

20-30 & $184 \pm 4$ &  $1.41 \pm 0.08$ & $0.85 \pm 0.03$& $166 \pm 3$ & $1.66 \pm 0.09$ & $0.97 \pm     0.03  $     \\      
\hline 

30-40 & $186 \pm 4$ & $1.30 \pm 0.08$ & $0.83 \pm 0.03$& $164 \pm 3$ & $1.62 \pm 0.09$ & $0.97 \pm  0.03 $    \\      
\hline  

40-50 & $189 \pm 4$ & $1.20 \pm 0.07$ & $0.80 \pm 0.02$ & $161 \pm 3$ & $1.60 \pm 0.09$ & $0.96 \pm 0.03$       \\      
\hline    

50-70 & $193 \pm 6$ &  $1.08 \pm 0.08$ & $0.76 \pm 0.02$& $155 \pm 3$ & $1.61 \pm 0.09$ & $0.96 \pm  0.03$     \\      
\hline     

70-100 & $194 \pm 8$ & $0.97 \pm 0.10$ & $0.70 \pm 0.03$    
& $144 \pm 3$ &  $1.68 \pm 0.08$ & $1.00 \pm 0.04$\\      
\hline     

\end{tabular}
\end{table*}

*/

In this work, first we have studied multiplicity dependence of different CFO schemes in p+p collisions at 
$\sqrt{s}$ = 7 TeV.
For this, we have used data on $\pt$ integrated hadron 
yield in p+p collisions measured at mid-rapidity at $\sqrt{s}$ = 7 TeV~\cite{Acharya:2018orn}. 
We have performed our analysis for grand canonical, canonical and strangeness canonical ensembles using a thermal model called THERMUS ~\cite{Wheaton:2004qb}. Further, for the grand canonical ensemble, three 
different freeze-out schemes- 1CFO, 1CFO$+\gamma_S$ and 2CFO are considered. For 1CFO and 1CFO$+\gamma_S$ schemes standard version of THERMUS is used and it is extended to include the 2CFO scheme.

To extract the CFO parameters we use available midrapidity yields of the following hadrons
($\pi^++\pi^-$)/2, ($K^++K^-$)/2, $K_s^0$, ($p+\bar{p}$)/2, ($K^{*0}+\bar{K^{*0}}$)/2, $\phi$,
($\Lambda+\bar{\Lambda}$)/2 and
($\Xi+\bar{\Xi}$)/2. Since an almost equal number of particle and antiparticle are produced
at this energy, chemical potentials are taken as zero for all the CFO schemes.
Therefore, the parameters to be extracted from the thermal fits in 1CFO are only the fireball volume 
$V$ and temperature $T$ at the chemical freeze-out. In the 1CFO$+\gamma_S$ scheme, we have one extra parameter
$\gamma_S$.
In 2CFO scheme, different freeze-out volume and temperature ($V_S$, $T_S$, $V_{NS}$ and $T_{NS}$) 
for the non-strange and strange hadrons are there. 
In this CFO scheme, the $\gamma_S$ has no role and hence it is regarded as 1. 
Therefore, the number of degrees of freedom (ndf = number of data points - number of free parameters)
in 1CFO, 1CFO+$\gamma_S$ and 2CFO schemes are 5,4 and 3 respectively. For canonical and strangeness canonical ensembles free parameters are $T, V$ and $\gamma_S$.
In Fig. \ref{fig:chisqr_pp} we compare the goodness of fit in terms of $\chi^2/$ndf 
of different CFO schemes in different multiplicity classes where the same is defined in terms of 
charge particle multiplicity.
Fitting quality is good for 1CFO+$\gamma_S$ at lowest $\langle dN_{ch}/d\eta \rangle$
where $\chi^2/$ndf is approximately 3. The $\chi^2/$ndf increases slightly with increasing $\langle dN_{ch}/d\eta \rangle$
and remains within the range 5-6 in other multiplicity classes. For the 1CFO scheme, $\chi^2/$ndf is approximately 8 at the 
lowest $\langle dN_{ch}/d\eta \rangle$ and reaches the highest value ($\sim 12$) 
when $\langle dN_{ch}/d\eta \rangle$ is within 3-6. The $\chi^2/$ndf then 
starts decreasing with increasing $\langle dN_{ch}/d\eta \rangle$. Compared to 1CFO+$\gamma_S$ scheme, 
the fitting quality is bad in 1CFO scheme in all the multiplicity classes.
Variation of $\chi^2/$ndf with $\langle dN_{ch}/d\eta \rangle$ for 2CFO scheme is almost similar to that of 1CFO.
The only difference is that fitting quality becomes comparable to that of 1CFO+$\gamma_S$ or even better 
at some multiplicity classes when $\langle dN_{ch}/d\eta \rangle > 10$.

\subsubsection{Results of canonical ensemble}
Along with the results of the grand canonical ensemble, we have also shown the quality of fitting in the canonical ensemble (CE) in Fig. \ref{fig:chisqr_pp}.
At the lowest $\langle dN_{ch}/d\eta \rangle$, $\chi^2/$ndf for CE is 15.59 which is much larger than that of three CFO schemes in grand canonical ensembles. It then decreases with increase of $\langle dN_{ch}/d\eta \rangle$. The $\chi^2/$ndf for CE is greater than that of 1CFO+$\gamma_s$ scheme in the whole range of multiplicities shown in this figure. Another significant difference is that the extracted chemical freeze-out temperature in CE is too large compared to those of GCE and QCD crossover temperature which is around 155 MeV. At the lowest multiplicity, $T$ is around 194 MeV. Temperature decreases with increasing multiplicity and it reaches 173 MeV at the highest multiplicity. 
We have observed that  $\phi$ meson is responsible for large $\chi^2/$ndf in CE. 
If we exclude $\phi$ in the fitting, $\chi^2/$ndf in the lowest multiplicity class goes down from 15.59 to 2.54.
We compared our result of CE with the previous work in Ref. \cite{Sharma:2018jqf}. It has been shown in Ref. \cite{Sharma:2018jqf} that CE gives the best result in the low multiplicity region which seems 
to contradict our result. However, this is not the case as different set of particles are used in both analysis. Particularly in Ref. \cite{Sharma:2018jqf} $K_s, K^{*}$ and $\phi$ are excluded from the fitting. 
Another difference is the inclusion of $\Omega$ which is not there in our fitting. 
Already we have discussed the issue with the $\phi$ meson. 
Further we checked the effect on our results by excluding $K_s, K^{*}$ and $\phi$. If we remove only $K^{*}$ and $\phi$, the $\chi^2/$ndf at lowest $\langle dN_{ch}/d\eta \rangle$ goes down from 15.59 to 1.28.
By excluding $K^{*}$, $\phi$ and $K_s$ the $\chi^2/$ndf becomes even better (0.9). Corresponding chemical freezeout temperatures in these two cases are 201 and 202 MeV respectively. So, although there is an improvement of $\chi^2/$ndf chemical freeze-out temperature is still absurd. This indicates that the CE is not suitable for LHC energy. In presence of $\phi$ meson, a large freezeout temperature in CE was also reported in Ref. \cite{Vovchenko:2019kes}. The reason for the failure of CE is probably because the accepted rapidity region which is much smaller than the total rapidity. The system is more like a grand canonical rather than canonical ensemble. There is no reason that the baryon, strangeness and electric charge will be exactly conserved at one unit of rapidity region. It is argued in Ref. \cite{Vovchenko:2019kes} for the exact conservation of these charges several units of rapidity interval around the midrapidity is needed rather than a single unit. Because of these problems in CE we have not considered CE in the rest of the paper.
 
\subsubsection{Results of strangeness canonical ensemble}
In Fig.\ref{fig:chisqr_pp}, the worst fitting is observed in case of SCE. The $\chi^2/$ndf is $\sim 20$ at the lowest $\langle dN_{ch}/d\eta \rangle$. Similar to CE, $\chi^2/$ndf for SCE decreases with increase of $\langle dN_{ch}/d\eta \rangle$. The $\chi^2/$ndf for SCE is always larger than that of 1CFO+$\gamma_s$ scheme. Again a relatively better $\chi^2/$ndf is shown in Ref. \cite{Sharma:2018jqf} but with the cost of giving up  $K_s, K^{*}$ and $\phi$ in the fitting. 
Since $\chi^2/$ndf in Fig.\ref{fig:chisqr_pp} is bad in SCE, we will not 
consider this CFO scheme in the rest of the paper.

Overall from Fig. \ref{fig:chisqr_pp}, we can say that, except a few multiplicity classes,
1CFO+$\gamma_S$ scheme is better compared to other CFO schemes.

\subsubsection{Extracted chemical freeze-out parameters and comparison between data and model}
\begin{figure}[!hbtb]
 \begin{center}
\includegraphics[scale=0.4]{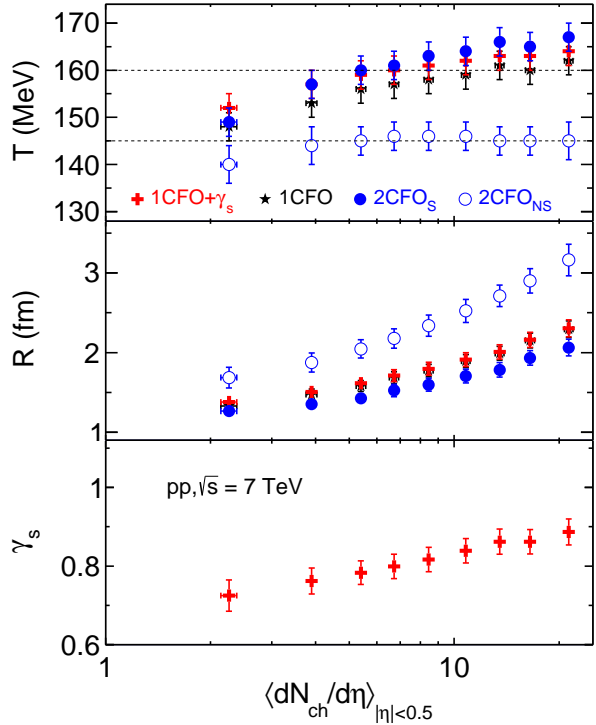}
\caption{(Color online) The multiplicity dependence of chemical freeze-out parameters (temperature $T$, radius $R$ and 
 strangeness non-equilibrium factor $\gamma_S$) obtained in three different freeze-out schemes.}
 \label{fig:parpp7TeV}
 \end{center}
\end{figure}

Extracted chemical freeze-out parameters in different multiplicity classes are listed in the table-\ref{tab:fitparameter} 
where instead of $\langle dN_{ch}/d\eta \rangle$, percentage of centrality is mentioned~\cite{Acharya:2018orn}.
Note that the $\langle dN_{ch}/d\eta \rangle$ decreases as the percentage of the centrality increases.
The $\langle dN_{ch}/d\eta \rangle$ dependence of chemical freeze-out
temperature, the radius of the fireball and the strangeness suppression factor in three different CFO schemes are shown in 
Fig.\ref{fig:parpp7TeV}. At the lowest value of $\langle dN_{ch}/d\eta \rangle$ temperature is around 152 MeV for 1CFO+$\gamma_S$ scheme. The 
temperature gradually increases with the increase of $\langle dN_{ch}/d\eta \rangle$ and reaches around 164 MeV at the highest $\langle dN_{ch}/d\eta \rangle$.
Similar temperatures are obtained for the 1CFO scheme as well. On the other hand for the 2CFO scheme,
a clear separation of freeze-out temperature of the non-strange and strange
hadrons is observed and the separation increases with increasing $\langle dN_{ch}/d\eta \rangle$.
The multiplicity dependence of freeze-out temperature of strange hadrons in the 2CFO scheme is
similar to the 1CFO and 1CFO+$\gamma_S$ schemes. 
However, no such multiplicity dependence of freeze-out temperature is observed for the non-strange hadrons in this CFO scheme.
Not only that, freeze-out temperature of non-strange hadrons, which lies around 145 MeV, 
is significantly (5-20 MeV) lower compared to that of strange hadrons which
suggests that the strange hadrons freeze earlier compared to non-strange hadrons. Compared to different CFO schemes in the grand canonical ensemble, the larger temperature is observed for the canonical ensemble in all $\langle dN_{ch}/d\eta \rangle$.
Although it is not shown still we would like to mention here that the extracted temperature at lowest $\langle dN_{ch}/d\eta \rangle$ for CE is around 195 MeV which is very large compared to other CFO schemes and therefore seems unrealistic.

The middle panel of Fig.\ref{fig:parpp7TeV} shows the multiplicity dependence
of fireball radius at the freeze-out. As expected, $R$ increases with increasing $\langle dN_{ch}/d\eta \rangle$ in all the CFO schemes in the grand canonical ensemble.
Similar to the temperature, freeze-out radii are also close to each other in 1CFO and 1CFO+$\gamma_S$ schemes.
In these two CFO schemes, $R$ is around 1.3 fm at the lower $\langle dN_{ch}/d\eta \rangle$ and reaches upto $\sim$2.3 fm at the highest $\langle dN_{ch}/d\eta \rangle$.
Again in the 2CFO scheme, two separate freeze-out radii is observed for strange and non-strange hadrons.
Extracted size of the fireball for non-strange hadrons is larger than that of strange hadrons in all multiplicity classes.

The multiplicity dependence of the strangeness suppression factor is shown in the lower panel of Fig.\ref{fig:parpp7TeV}.
At the lowest  $\langle dN_{ch}/d\eta \rangle$, the $\gamma_S$ is around 0.72 which is far away from the equilibrium value 1.
With the increase of $\langle dN_{ch}/d\eta \rangle$, the $\gamma_S$ gradually increases towards its equilibrium value
and reaches up to 0.89 at highest $\langle dN_{ch}/d\eta \rangle$.
It indicates that full strangeness equilibration is not achieved at lower values of $\langle dN_{ch}/d\eta \rangle$. 
However, it is
moving towards the equilibrium when $\langle dN_{ch}/d\eta \rangle$ is relatively larger. 
Therefore, it is expected that the CFO scheme which deals with the non-equilibrium situation of the strange hadron
is the best suited when $\langle dN_{ch}/d\eta \rangle$ is small. 
For this reason, in the previous figure (Fig. \ref{fig:chisqr_pp}), better fitting quality 
(i.e, lower $\chi^2$/ndf) 
at lower $\langle dN_{ch}/d\eta \rangle$ was observed in 1CFO+$\gamma_S$ scheme compared to other 1CFO and 2CFO schemes where full equilibration was assumed.

We have observed that the extracted freeze-out temperature and $\gamma_s$ in 1CFO+$\gamma_s$ scheme in this present work is in good agreement with those of Ref. \cite{Sharma:2018jqf} despite having the difference in the fitted particle list mentioned previously. 

\begin{figure*}[!ht]
 \begin{center}
 \includegraphics[scale=0.8]{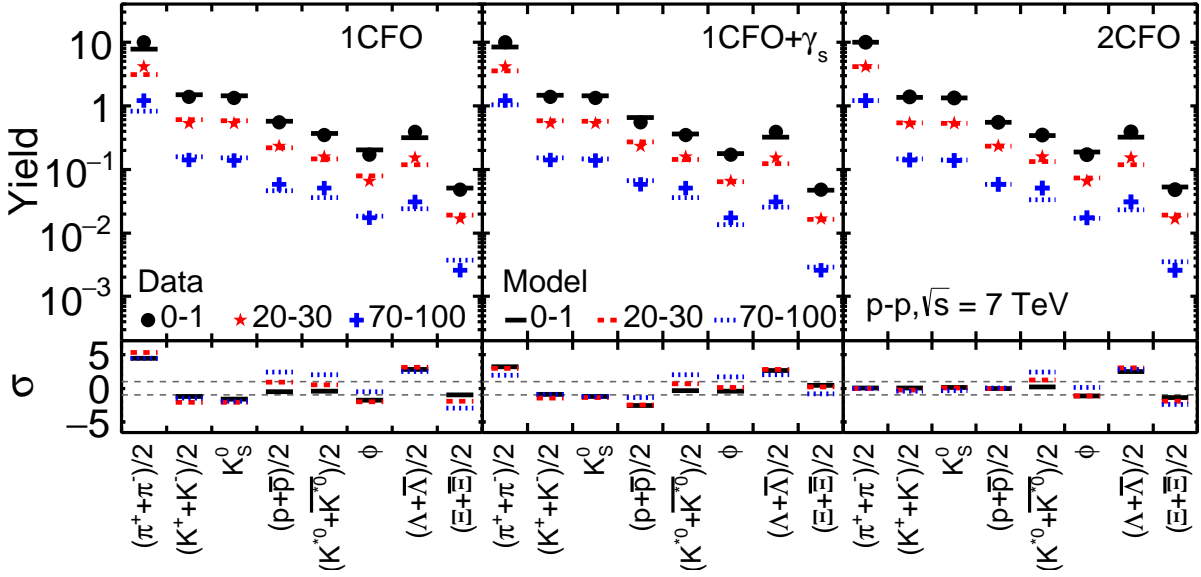}
 \caption{(Color online) The comparison between data~\cite{Acharya:2018orn} and model~\cite{Wheaton:2004qb}
 for p+p collisions at $\sqrt{s}$ = 7 TeV in three different freeze-out schemes: 
 1CFO (left column), 1CFO$+\gamma_S$  (middle column) and 2CFO (right column). 
 The deviation of the data from the model (Eq.~\ref{eq:dev}) for each freeze-out scheme are also shown.}
 \label{fig:spectra}	
 \end{center}
\end{figure*}

Now we will discuss the comparison between experimental data of hadrons and model estimation. For the rest of the paper, we will consider CFO schemes in grand canonical ensemble only since in Fig. \ref{fig:chisqr_pp} we have already seen that fitting quality is much better in grand canonical ensemble combined to CE or SCE.
In the upper panels of Fig. \ref{fig:spectra}, we show the comparison
between experimentally measured hadronic yields and the 
best-fitted model calculation in three CFO schemes: 1CFO (left column), 1CFO$+\gamma_S$  (middle column) and 
2CFO (right column).
For the illustration purpose, we choose only three multiplicity
bins: 0-1 \%, 20-30 \% and 70-100 \%. Experimental data is indicated by solid points while a line symbol is used to show the model estimation.
The lower panel of this figure shows the deviation which is defined as
\begin{equation}\label{eq:dev}
 \sigma = \frac{\text{Data - Model}}{\text{Error of Data}}.
\end{equation}
Horizontal lines at $\pm 1$ are drawn to indicate $\pm 1 \sigma$ deviation.
Deviations are less than $3\sigma$ for all the hadrons except pion in the 1CFO scheme.
In 1CFO+$\gamma_S$ scheme, significant improvement is observed particularly in the 70-100 \% centrality class where the modulus of
deviation is less than $2\sigma$ for all the hadrons including pion. On the other hand, for all the centrality classes
descriptions of $\pi, K, K^{0}_{s}$ and $p$ are very good (modulus of the deviation is less than 1$\sigma$) in the 2CFO scheme. 
Although, it is not good for strange hadrons like $K^{\star}, \Lambda$ and $\Xi$ in 70-100 \% centrality class.
These hadrons are described better in the 1CFO+$\gamma_S$ scheme.
Note that a separate freeze-out surface is used for the hadrons containing the strange quark in the 2CFO scheme;
still, its performance is not better than the 1CFO+$\gamma_S$ scheme.
The description of both non-strange and strange hadrons
in 70-100 \% centrality is better in 1CFO+$\gamma_S$ scheme 
compared to the other two CFO schemes. 
The $\chi^2/$ndf is also smaller in 1CFO+$\gamma_S$ scheme (see the Fig. \ref{fig:chisqr_pp}).
Therefore, we can say that 1CFO+$\gamma_S$ is the most 
appropriate freeze-out scenario for 70-100 \% centrality class.
The same is true for 20-30\% multiplicity classes as well. As we move toward higher multiplicity class, improvement is observed in 2CFO scheme. In 0-1\% multiplicity class, description of $\pi, K, K^{0}_{s}$ and $p$ is better in 2CFO scheme compared to 1CFO and 1CFO+$\gamma_S$ schemes. As a result, $\chi^2/$ndf is slightly small in case of 2CFO scheme compared to other two CFO schemes in this multiplicity class.

\subsection{Comparison of different freeze-out scenarios at LHC}

\begin{figure}[!hbtb]
 \begin{center}
 \includegraphics[scale=0.4]{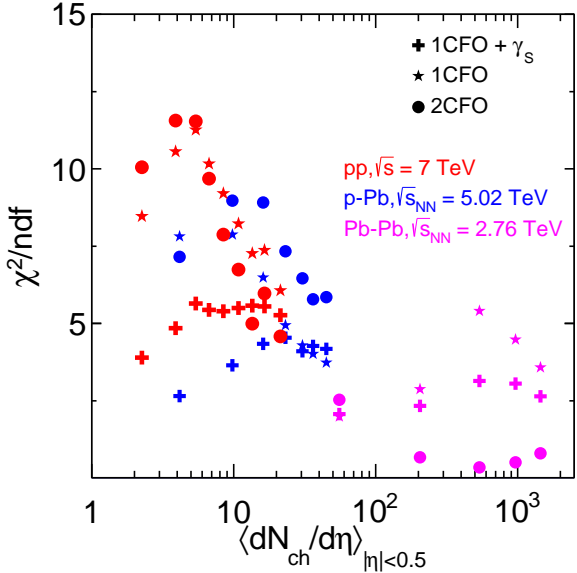}
 \caption{(Color online)
 $\chi^{2}/ndf$ in three different CFO schemes in grand canonical ensemble as a function of average charged particle multiplicity in p+p collisions 
 at $\sqrt{s} = 7$ TeV and compared with the same in p+Pb and Pb+Pb collisions at $\sNN$ = 5.02 
 \cite{Abelev:2013haa,Adam:2015vsf,Adam:2016bpr} and 2.76  TeV 
 \cite{Abelev:2013vea,Abelev:2013xaa,ABELEV:2013zaa,Abelev:2014uua}, respectively. }
 \label{fig:chi2LHC}	
 \end{center}
\end{figure}

\begin{figure}[!hbtb]
\begin{center}
\includegraphics[scale=0.377]{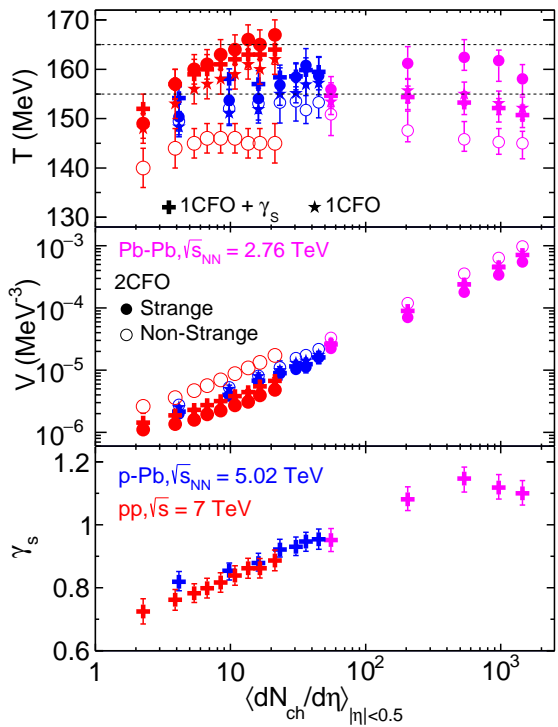}
\caption{(Color online)
Extracted freeze-out parameters in three different CFO schemes in grand canonical ensemble as a function  $\langle dN_{ch}/d\eta \rangle$ considering three different collision system p+p at $\sqrt{s}$ = 7 TeV, p+Pb at $\sNN$ = 5.02 and Pb+Pb collisions at $\sNN$ = 2.76 TeV. 
}
 \label{fig:cfoLHC}
 \end{center}
\end{figure}

\begin{figure}[!hbtb]
 \begin{center}
 \includegraphics[scale=0.4]{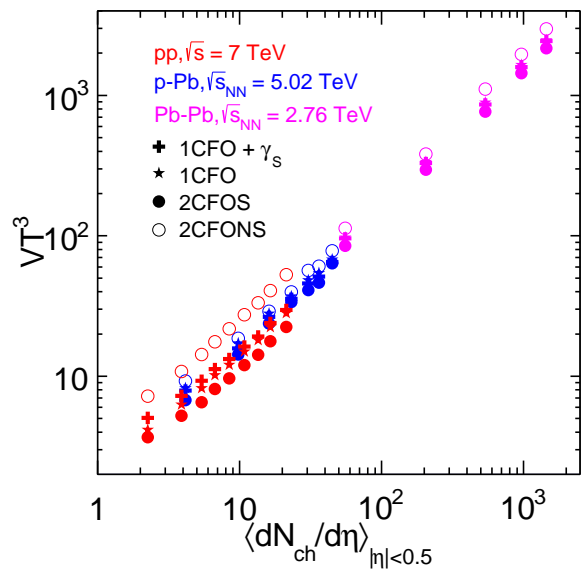}
 \caption{(Color online) Product of extracted volume and temperature cube as a function of average charged particle
 multiplicity in p+p collisions 
 at $\sqrt{s} = 7$ TeV and compared with the same in p+Pb and Pb+Pb collisions at $\sNN$ = 5.02 and 2.76  
 TeV \cite{Chatterjee:2016cog}, respectively for the best 
 freeze-out scheme.}
 \label{fig:VolLHC}	
 \end{center}
\end{figure}

We will now present a broader picture of the multiplicity dependence of freeze-out scenarios at the LHC.
Here we will consider collision systems p+Pb and Pb+Pb along with p+p.
In Fig.\ref{fig:chi2LHC}, we have shown the multiplicity dependence of least $\chi^2$/ndf. 
Here we have taken some data from the previously published results of p+Pb and Pb+Pb collisions at $\sNN$ = 5.02 and 2.76 TeV, respectively \cite{Chatterjee:2016cog}
combined with the present results of p+p collision at $\sqrt{s} = 7$ TeV.
The left panel shows
 $\chi^{2}/ndf$ in three different CFO schemes in grand canonical ensemble as a function of $\langle dN_{ch}/d\eta \rangle$. Different marker style is used for different CFO schemes, while different colour is used for different collision system. For the p+p collision, the result is already discussed in Fig. \ref{fig:chisqr_pp}. For p+Pb collision,
lowest $\chi^2$/ndf is observed in 1CFO+$\gamma_s$ scheme when $dN_{ch}/d\eta \rangle < 30$. Above this 1CFO scheme performs better. While for Pb+Pb collision, 2CFO gives best result when $\langle dN_{ch}/d\eta \rangle > 100$. Below this multiplicity, 1CFO performs well. 
The variation of $\langle dN_{ch}/d\eta \rangle$ at LHC is more than three orders of magnitude.
Based on the applicability of different CFO schemes,
we can divide the whole range of $\langle dN_{ch}/d\eta \rangle$ into
three regions: low ($\langle dN_{ch}/d\eta \rangle < 30$), intermediate ($30 < \langle dN_{ch}/d\eta \rangle < 60$)
and high ($\langle dN_{ch}/d\eta \rangle > 100$) multiplicities. 
Note that presently no data is available within $60 < \langle dN_{ch}/d\eta \rangle < 100$.
All p+p collisions and p+Pb collisions where $\langle dN_{ch}/d\eta \rangle < 30$ are best described
by 1CFO+$\gamma_S$ scheme which is indicated by plus marker.
It indicates that strange hadrons containing strange quark are out of equilibrium when $\langle dN_{ch}/d\eta \rangle < 30$.
In small systems, the fireball lifetime is expected to be small. 
As a result, interaction among constituents is not sufficient enough to achieve equilibration.
The equilibrium CFO schemes are good only if $\langle dN_{ch}/d\eta \rangle$ is approximately greater than 30.
For the intermediate range of multiplicity ($30 < \langle dN_{ch}/d\eta \rangle < 60$)
1CFO scheme has least $\chi^2$/ndf which is shown by star marker. 
High multiplicity p+Pb and low multiplicity Pb+Pb collisions fall in this category.
The Pb+Pb collisions with $\langle dN_{ch}/d\eta \rangle > 100$ can be best described by the 2CFO scheme which is shown by a circular marker.

In three panels of Fig.\ref{fig:cfoLHC} we show the multiplicity dependence of  freeze-out temperature (top panel), the volume of the 
fireball (middle panel) and the $\gamma_S$ (bottom panel) at the LHC using three different CFO schemes in grand canonical ensemble. 
freeze-out parameters of p+Pb and Pb+Pb collisions at $\sNN$ = 5.02 and 2.76 TeV, respectively are taken from \cite{Chatterjee:2016cog}.
We have already discussed the results of p+p collision at $\sqrt{s}$ = 7 TeV in Fig. \ref{fig:parpp7TeV}.
One can see that compared to  p+p collision the freeze-out temperatures in different CFO schemes for p+Pb collision are much closer to each other and lie around 155 MeV. Further, no multiplicity dependence of temperature is observed in the case of p+Pb collision. On the other hand for Pb+Pb collision, the freeze-out temperature is around 155 MeV for 1CFO and 1CFO+$\gamma_s$ schemes while for the 2CFO scheme, the temperature is slightly higher than 155 MeV for strange hadrons and slightly lower for non-strange hadrons. Here also no significant multiplicity dependence is observed.
Now let us discuss the best-fitted results of freeze-out temperature.
For the low and intermediate regions of multiplicity ($\langle dN_{ch}/d\eta \rangle < 60$),
best-fitted (for which $\chi^{2}/ndf$ is minimum) freeze-out temperature lies within 155-165 MeV. 
In the high multiplicity region ($\langle dN_{ch}/d\eta \rangle > 100$) freeze-out temperature of the hadrons containing strange quark also
lies within this range. However, for non-strange hadrons, the temperature is approximately 15 MeV less.
It means non-strange hadrons freeze-out later than the strange hadrons when $\langle dN_{ch}/d\eta \rangle > 100$.
The volume of the fireball at the freeze-out increases almost linearly with the multiplicity. 
Volumes for 1CFO and 1CFO+$\gamma_s$ are almost same in all $\langle dN_{ch}/d\eta \rangle$.
In the 2CFO scheme, freeze-out volume of strange and non-strange are different and it is larger for the non-strange hadrons
since they freeze-out later. Overall we observe a smooth variation of volume from one collision system to another.
With the increase of multiplicity, the $\gamma_S$ increases towards unity and for Pb+Pb collision $\gamma_S$ is greater than one.
Values of $\gamma_S$ for different data sets 
lie in a line. Here also we observe a smooth variation from one system to another. 
Note the best fit result suggest that the $\gamma_S$ is only needed for the low multiplicity events. Above $\langle dN_{ch}/d\eta \rangle \approx 30$
an equilibration CFO schemes, either 1CFO or 2CFO can be used.

Instead of multiplicity, we can separate three zones in terms of dimension-less quality $VT^3$ as well
since it is proportional to the $\langle dN_{ch}/d\eta \rangle$ and this is presented in Fig. \ref{fig:VolLHC}. 
This figure shows that the 1CFO+$\gamma_S$ scheme is the best for the small freeze-out system where $VT^3 <50$.
For intermediate system size, where $50 < VT^3 < 100$,  1CFO performs well whereas for a large system, where $VT^3 > 100$, 
the 2CFO scheme is more suitable. We observe that the best fit value of $VT^3$ increases almost linearly with the multiplicity 
even though different CFO scheme is suitable for the different zone of multiplicity. Further, Fig.\ref{fig:VolLHC} shows that our 
treatment in the grand canonical ensemble is justified since $VT^3>1$ in all the multiplicity classes for all the collision systems~\cite{Kraus:2008fh} 

\section{Summary}
\label{sum}

We have studied the transverse momentum integrated hadron yields in different multiplicity classes of p+p collisions at 
$\sqrt{s} = 7$ TeV using a thermal model (THERMUS) and extracted the CFO parameters. Analysis has been done for 
different CFO schemes $e.g.,$ 1CFO, 1CFO+$\gamma_S$, 2CFO in the grand canonical ensemble. The analysis is 
also done for canonical and strangeness canonical ensemble. The charge conservation is enforced across only a single 
unit of rapidity within which experimental data are available. Within this limitation it has been observed that for most of the 
multiplicity classes 1CFO+$\gamma_S$ scheme best describe the data. The value of $\gamma_S$ varies between 0.72 to 0.89
which indicates that hadrons containing strange quark are out-of-equilibrium at the freeze-out in all multiplicity classes. 
With the increase of multiplicity the chemical freeze-out temperature increases from 152 MeV to 164 MeV. The fireball size 
also increases with increases of multiplicity and the radius varies between 1.38 - 2.3 fm.

We have also tried to understand the applicability of different CFO schemes at LHC considering two other 
collision systems p+Pb and Pb+Pb at $\sNN = $ 5.02 and 2.76 TeV respectively. We observed a multiplicity 
(or freeze-out volume) dependence of CFO schemes instead of colliding ion and energy dependence. The 
1CFO+$\gamma_S$ scheme is suitable to describe the hadronic yield when multiplicity is less than 30 whereas 
2CFO perform the best when multiplicity is larger than 100. In the intermediate region $30 < dN_{ch}/d\eta < 60$ 
1CFO gives the best description of the data. The same conclusion can be drawn in terms of freeze-out volume as 
well. The 1CFO+$\gamma_S$, 1CFO and 2CFO schemes best description of the data when the dimensionless 
quantity $VT^3$ satisfies the conditions $VT^3 <50$, $50 < VT^3 < 100$ and $VT^3 > 100$ respectively.

\section{Acknowledgement}
AKD and RS acknowledge the support of XII th plan project no. 12-R$\&$D-NIS-5.11-0300 of Govt. of India.
BM acknowledges financial support from J C Bose National Fellowship of DST Govt. of India.
SS acknowledges financial support from the Polish National Agency for Academic Exchange 
through Ulam Scholarship with the agreement
no: PPN/ULM/2019/1/00093/U/00001.
\bibliographystyle{apsrev4-1}
\bibliography{ppThermus.bib}

\begin{thebibliography}{27}%
\makeatletter
\providecommand \@ifxundefined [1]{%
 \@ifx{#1\undefined}
}%
\providecommand \@ifnum [1]{%
 \ifnum #1\expandafter \@firstoftwo
 \else \expandafter \@secondoftwo
 \fi
}%
\providecommand \@ifx [1]{%
 \ifx #1\expandafter \@firstoftwo
 \else \expandafter \@secondoftwo
 \fi
}%
\providecommand \natexlab [1]{#1}%
\providecommand \enquote  [1]{``#1''}%
\providecommand \bibnamefont  [1]{#1}%
\providecommand \bibfnamefont [1]{#1}%
\providecommand \citenamefont [1]{#1}%
\providecommand \href@noop [0]{\@secondoftwo}%
\providecommand \href [0]{\begingroup \@sanitize@url \@href}%
\providecommand \@href[1]{\@@startlink{#1}\@@href}%
\providecommand \@@href[1]{\endgroup#1\@@endlink}%
\providecommand \@sanitize@url [0]{\catcode `\\12\catcode `\$12\catcode
  `\&12\catcode `\#12\catcode `\^12\catcode `\_12\catcode `\%12\relax}%
\providecommand \@@startlink[1]{}%
\providecommand \@@endlink[0]{}%
\providecommand \url  [0]{\begingroup\@sanitize@url \@url }%
\providecommand \@url [1]{\endgroup\@href {#1}{\urlprefix }}%
\providecommand \urlprefix  [0]{URL }%
\providecommand \Eprint [0]{\href }%
\providecommand \doibase [0]{http://dx.doi.org/}%
\providecommand \selectlanguage [0]{\@gobble}%
\providecommand \bibinfo  [0]{\@secondoftwo}%
\providecommand \bibfield  [0]{\@secondoftwo}%
\providecommand \translation [1]{[#1]}%
\providecommand \BibitemOpen [0]{}%
\providecommand \bibitemStop [0]{}%
\providecommand \bibitemNoStop [0]{.\EOS\space}%
\providecommand \EOS [0]{\spacefactor3000\relax}%
\providecommand \BibitemShut  [1]{\csname bibitem#1\endcsname}%
\let\auto@bib@innerbib\@empty
\bibitem [{\citenamefont {Braun-Munzinger}\ \emph {et~al.}(1996)\citenamefont
  {Braun-Munzinger}, \citenamefont {Stachel}, \citenamefont {Wessels},\ and\
  \citenamefont {Xu}}]{BraunMunzinger:1995bp}%
  \BibitemOpen
  \bibfield  {author} {\bibinfo {author} {\bibfnamefont {P.}~\bibnamefont
  {Braun-Munzinger}}, \bibinfo {author} {\bibfnamefont {J.}~\bibnamefont
  {Stachel}}, \bibinfo {author} {\bibfnamefont {J.~P.}\ \bibnamefont
  {Wessels}}, \ and\ \bibinfo {author} {\bibfnamefont {N.}~\bibnamefont {Xu}},\
  }\href {\doibase 10.1016/0370-2693(95)01258-3} {\bibfield  {journal}
  {\bibinfo  {journal} {Phys. Lett.}\ }\textbf {\bibinfo {volume} {B365}},\
  \bibinfo {pages} {1} (\bibinfo {year} {1996})},\ \Eprint
  {http://arxiv.org/abs/nucl-th/9508020} {arXiv:nucl-th/9508020 [nucl-th]}
  \BibitemShut {NoStop}%
\bibitem [{\citenamefont {Yen}\ and\ \citenamefont
  {Gorenstein}(1999)}]{Yen:1998pa}%
  \BibitemOpen
  \bibfield  {author} {\bibinfo {author} {\bibfnamefont {G.~D.}\ \bibnamefont
  {Yen}}\ and\ \bibinfo {author} {\bibfnamefont {M.~I.}\ \bibnamefont
  {Gorenstein}},\ }\href {\doibase 10.1103/PhysRevC.59.2788} {\bibfield
  {journal} {\bibinfo  {journal} {Phys. Rev.}\ }\textbf {\bibinfo {volume}
  {C59}},\ \bibinfo {pages} {2788} (\bibinfo {year} {1999})},\ \Eprint
  {http://arxiv.org/abs/nucl-th/9808012} {arXiv:nucl-th/9808012 [nucl-th]}
  \BibitemShut {NoStop}%
\bibitem [{\citenamefont {Braun-Munzinger}\ \emph {et~al.}(1999)\citenamefont
  {Braun-Munzinger}, \citenamefont {Heppe},\ and\ \citenamefont
  {Stachel}}]{BraunMunzinger:1999qy}%
  \BibitemOpen
  \bibfield  {author} {\bibinfo {author} {\bibfnamefont {P.}~\bibnamefont
  {Braun-Munzinger}}, \bibinfo {author} {\bibfnamefont {I.}~\bibnamefont
  {Heppe}}, \ and\ \bibinfo {author} {\bibfnamefont {J.}~\bibnamefont
  {Stachel}},\ }\href {\doibase 10.1016/S0370-2693(99)01076-X} {\bibfield
  {journal} {\bibinfo  {journal} {Phys. Lett.}\ }\textbf {\bibinfo {volume}
  {B465}},\ \bibinfo {pages} {15} (\bibinfo {year} {1999})},\ \Eprint
  {http://arxiv.org/abs/nucl-th/9903010} {arXiv:nucl-th/9903010 [nucl-th]}
  \BibitemShut {NoStop}%
\bibitem [{\citenamefont {Andronic}\ \emph {et~al.}(2006)\citenamefont
  {Andronic}, \citenamefont {Braun-Munzinger},\ and\ \citenamefont
  {Stachel}}]{Andronic:2005yp}%
  \BibitemOpen
  \bibfield  {author} {\bibinfo {author} {\bibfnamefont {A.}~\bibnamefont
  {Andronic}}, \bibinfo {author} {\bibfnamefont {P.}~\bibnamefont
  {Braun-Munzinger}}, \ and\ \bibinfo {author} {\bibfnamefont {J.}~\bibnamefont
  {Stachel}},\ }\href {\doibase 10.1016/j.nuclphysa.2006.03.012} {\bibfield
  {journal} {\bibinfo  {journal} {Nucl. Phys.}\ }\textbf {\bibinfo {volume}
  {A772}},\ \bibinfo {pages} {167} (\bibinfo {year} {2006})},\ \Eprint
  {http://arxiv.org/abs/nucl-th/0511071} {arXiv:nucl-th/0511071 [nucl-th]}
  \BibitemShut {NoStop}%
\bibitem [{\citenamefont {Becattini}(1996)}]{Becattini:1995if}%
  \BibitemOpen
  \bibfield  {author} {\bibinfo {author} {\bibfnamefont {F.}~\bibnamefont
  {Becattini}},\ }\href {\doibase 10.1007/BF02907431} {\bibfield  {journal}
  {\bibinfo  {journal} {Z. Phys.}\ }\textbf {\bibinfo {volume} {C69}},\
  \bibinfo {pages} {485} (\bibinfo {year} {1996})}\BibitemShut {NoStop}%
\bibitem [{\citenamefont {Becattini}\ \emph {et~al.}(1996)\citenamefont
  {Becattini}, \citenamefont {Giovannini},\ and\ \citenamefont
  {Lupia}}]{Becattini:1995xt}%
  \BibitemOpen
  \bibfield  {author} {\bibinfo {author} {\bibfnamefont {F.}~\bibnamefont
  {Becattini}}, \bibinfo {author} {\bibfnamefont {A.}~\bibnamefont
  {Giovannini}}, \ and\ \bibinfo {author} {\bibfnamefont {S.}~\bibnamefont
  {Lupia}},\ }\href {\doibase 10.1007/s002880050269, 10.1007/BF02909178}
  {\bibfield  {journal} {\bibinfo  {journal} {Z. Phys.}\ }\textbf {\bibinfo
  {volume} {C72}},\ \bibinfo {pages} {491} (\bibinfo {year} {1996})},\ \Eprint
  {http://arxiv.org/abs/hep-ph/9511203} {arXiv:hep-ph/9511203 [hep-ph]}
  \BibitemShut {NoStop}%
\bibitem [{\citenamefont {Becattini}\ and\ \citenamefont
  {Heinz}(1997)}]{Becattini:1997rv}%
  \BibitemOpen
  \bibfield  {author} {\bibinfo {author} {\bibfnamefont {F.}~\bibnamefont
  {Becattini}}\ and\ \bibinfo {author} {\bibfnamefont {U.~W.}\ \bibnamefont
  {Heinz}},\ }\href {\doibase 10.1007/s002880050551} {\bibfield  {journal}
  {\bibinfo  {journal} {Z. Phys.}\ }\textbf {\bibinfo {volume} {C76}},\
  \bibinfo {pages} {269} (\bibinfo {year} {1997})},\ \bibinfo {note} {[Erratum:
  Z. Phys.C76,578(1997)]},\ \Eprint {http://arxiv.org/abs/hep-ph/9702274}
  {arXiv:hep-ph/9702274 [hep-ph]} \BibitemShut {NoStop}%
\bibitem [{\citenamefont {Abelev}\ \emph
  {et~al.}(2012{\natexlab{a}})\citenamefont {Abelev} \emph
  {et~al.}}]{Abelev:2012wca}%
  \BibitemOpen
  \bibfield  {author} {\bibinfo {author} {\bibfnamefont {B.}~\bibnamefont
  {Abelev}} \emph {et~al.} (\bibinfo {collaboration} {ALICE}),\ }\href
  {\doibase 10.1103/PhysRevLett.109.252301} {\bibfield  {journal} {\bibinfo
  {journal} {Phys. Rev. Lett.}\ }\textbf {\bibinfo {volume} {109}},\ \bibinfo
  {pages} {252301} (\bibinfo {year} {2012}{\natexlab{a}})},\ \Eprint
  {http://arxiv.org/abs/1208.1974} {arXiv:1208.1974 [hep-ex]} \BibitemShut
  {NoStop}%
\bibitem [{\citenamefont {Bellwied}\ \emph {et~al.}(2013)\citenamefont
  {Bellwied}, \citenamefont {Borsanyi}, \citenamefont {Fodor}, \citenamefont
  {Katz},\ and\ \citenamefont {Ratti}}]{Bellwied:2013cta}%
  \BibitemOpen
  \bibfield  {author} {\bibinfo {author} {\bibfnamefont {R.}~\bibnamefont
  {Bellwied}}, \bibinfo {author} {\bibfnamefont {S.}~\bibnamefont {Borsanyi}},
  \bibinfo {author} {\bibfnamefont {Z.}~\bibnamefont {Fodor}}, \bibinfo
  {author} {\bibfnamefont {S.~D.}\ \bibnamefont {Katz}}, \ and\ \bibinfo
  {author} {\bibfnamefont {C.}~\bibnamefont {Ratti}},\ }\href {\doibase
  10.1103/PhysRevLett.111.202302} {\bibfield  {journal} {\bibinfo  {journal}
  {Phys. Rev. Lett.}\ }\textbf {\bibinfo {volume} {111}},\ \bibinfo {pages}
  {202302} (\bibinfo {year} {2013})},\ \Eprint {http://arxiv.org/abs/1305.6297}
  {arXiv:1305.6297 [hep-lat]} \BibitemShut {NoStop}%
\bibitem [{\citenamefont {Chatterjee}\ \emph {et~al.}(2013)\citenamefont
  {Chatterjee}, \citenamefont {Godbole},\ and\ \citenamefont
  {Gupta}}]{Chatterjee:2013yga}%
  \BibitemOpen
  \bibfield  {author} {\bibinfo {author} {\bibfnamefont {S.}~\bibnamefont
  {Chatterjee}}, \bibinfo {author} {\bibfnamefont {R.~M.}\ \bibnamefont
  {Godbole}}, \ and\ \bibinfo {author} {\bibfnamefont {S.}~\bibnamefont
  {Gupta}},\ }\href {\doibase 10.1016/j.physletb.2013.11.008} {\bibfield
  {journal} {\bibinfo  {journal} {Phys. Lett.}\ }\textbf {\bibinfo {volume}
  {B727}},\ \bibinfo {pages} {554} (\bibinfo {year} {2013})},\ \Eprint
  {http://arxiv.org/abs/1306.2006} {arXiv:1306.2006 [nucl-th]} \BibitemShut
  {NoStop}%
\bibitem [{\citenamefont {Chatterjee}\ \emph {et~al.}(2017)\citenamefont
  {Chatterjee}, \citenamefont {Dash},\ and\ \citenamefont
  {Mohanty}}]{Chatterjee:2016cog}%
  \BibitemOpen
  \bibfield  {author} {\bibinfo {author} {\bibfnamefont {S.}~\bibnamefont
  {Chatterjee}}, \bibinfo {author} {\bibfnamefont {A.~K.}\ \bibnamefont
  {Dash}}, \ and\ \bibinfo {author} {\bibfnamefont {B.}~\bibnamefont
  {Mohanty}},\ }\href {\doibase 10.1088/1361-6471/aa8857} {\bibfield  {journal}
  {\bibinfo  {journal} {J. Phys.}\ }\textbf {\bibinfo {volume} {G44}},\
  \bibinfo {pages} {105106} (\bibinfo {year} {2017})},\ \Eprint
  {http://arxiv.org/abs/1608.00643} {arXiv:1608.00643 [nucl-th]} \BibitemShut
  {NoStop}%
\bibitem [{\citenamefont {Abelev}\ \emph
  {et~al.}(2013{\natexlab{a}})\citenamefont {Abelev} \emph
  {et~al.}}]{Abelev:2013vea}%
  \BibitemOpen
  \bibfield  {author} {\bibinfo {author} {\bibfnamefont {B.}~\bibnamefont
  {Abelev}} \emph {et~al.} (\bibinfo {collaboration} {ALICE}),\ }\href
  {\doibase 10.1103/PhysRevC.88.044910} {\bibfield  {journal} {\bibinfo
  {journal} {Phys. Rev.}\ }\textbf {\bibinfo {volume} {C88}},\ \bibinfo {pages}
  {044910} (\bibinfo {year} {2013}{\natexlab{a}})},\ \Eprint
  {http://arxiv.org/abs/1303.0737} {arXiv:1303.0737 [hep-ex]} \BibitemShut
  {NoStop}%
\bibitem [{\citenamefont {Abelev}\ \emph
  {et~al.}(2013{\natexlab{b}})\citenamefont {Abelev} \emph
  {et~al.}}]{Abelev:2013xaa}%
  \BibitemOpen
  \bibfield  {author} {\bibinfo {author} {\bibfnamefont {B.~B.}\ \bibnamefont
  {Abelev}} \emph {et~al.} (\bibinfo {collaboration} {ALICE}),\ }\href
  {\doibase 10.1103/PhysRevLett.111.222301} {\bibfield  {journal} {\bibinfo
  {journal} {Phys. Rev. Lett.}\ }\textbf {\bibinfo {volume} {111}},\ \bibinfo
  {pages} {222301} (\bibinfo {year} {2013}{\natexlab{b}})},\ \Eprint
  {http://arxiv.org/abs/1307.5530} {arXiv:1307.5530 [nucl-ex]} \BibitemShut
  {NoStop}%
\bibitem [{\citenamefont {Abelev}\ \emph
  {et~al.}(2014{\natexlab{a}})\citenamefont {Abelev} \emph
  {et~al.}}]{ABELEV:2013zaa}%
  \BibitemOpen
  \bibfield  {author} {\bibinfo {author} {\bibfnamefont {B.~B.}\ \bibnamefont
  {Abelev}} \emph {et~al.} (\bibinfo {collaboration} {ALICE}),\ }\href
  {\doibase 10.1016/j.physletb.2014.05.052, 10.1016/j.physletb.2013.11.048}
  {\bibfield  {journal} {\bibinfo  {journal} {Phys. Lett.}\ }\textbf {\bibinfo
  {volume} {B728}},\ \bibinfo {pages} {216} (\bibinfo {year}
  {2014}{\natexlab{a}})},\ \bibinfo {note} {[Erratum: Phys.
  Lett.B734,409(2014)]},\ \Eprint {http://arxiv.org/abs/1307.5543}
  {arXiv:1307.5543 [nucl-ex]} \BibitemShut {NoStop}%
\bibitem [{\citenamefont {Abelev}\ \emph {et~al.}(2015)\citenamefont {Abelev}
  \emph {et~al.}}]{Abelev:2014uua}%
  \BibitemOpen
  \bibfield  {author} {\bibinfo {author} {\bibfnamefont {B.~B.}\ \bibnamefont
  {Abelev}} \emph {et~al.} (\bibinfo {collaboration} {ALICE}),\ }\href
  {\doibase 10.1103/PhysRevC.91.024609} {\bibfield  {journal} {\bibinfo
  {journal} {Phys. Rev.}\ }\textbf {\bibinfo {volume} {C91}},\ \bibinfo {pages}
  {024609} (\bibinfo {year} {2015})},\ \Eprint {http://arxiv.org/abs/1404.0495}
  {arXiv:1404.0495 [nucl-ex]} \BibitemShut {NoStop}%
\bibitem [{\citenamefont {Abelev}\ \emph
  {et~al.}(2012{\natexlab{b}})\citenamefont {Abelev} \emph
  {et~al.}}]{Abelev:2012hy}%
  \BibitemOpen
  \bibfield  {author} {\bibinfo {author} {\bibfnamefont {B.}~\bibnamefont
  {Abelev}} \emph {et~al.} (\bibinfo {collaboration} {ALICE}),\ }\href
  {\doibase 10.1140/epjc/s10052-012-2183-y} {\bibfield  {journal} {\bibinfo
  {journal} {Eur. Phys. J.}\ }\textbf {\bibinfo {volume} {C72}},\ \bibinfo
  {pages} {2183} (\bibinfo {year} {2012}{\natexlab{b}})},\ \Eprint
  {http://arxiv.org/abs/1208.5717} {arXiv:1208.5717 [hep-ex]} \BibitemShut
  {NoStop}%
\bibitem [{\citenamefont {Abelev}\ \emph
  {et~al.}(2012{\natexlab{c}})\citenamefont {Abelev} \emph
  {et~al.}}]{Abelev:2012jp}%
  \BibitemOpen
  \bibfield  {author} {\bibinfo {author} {\bibfnamefont {B.}~\bibnamefont
  {Abelev}} \emph {et~al.} (\bibinfo {collaboration} {ALICE}),\ }\href
  {\doibase 10.1016/j.physletb.2012.05.011} {\bibfield  {journal} {\bibinfo
  {journal} {Phys. Lett.}\ }\textbf {\bibinfo {volume} {B712}},\ \bibinfo
  {pages} {309} (\bibinfo {year} {2012}{\natexlab{c}})},\ \Eprint
  {http://arxiv.org/abs/1204.0282} {arXiv:1204.0282 [nucl-ex]} \BibitemShut
  {NoStop}%
\bibitem [{\citenamefont {Adam}\ \emph {et~al.}(2015)\citenamefont {Adam} \emph
  {et~al.}}]{Adam:2015qaa}%
  \BibitemOpen
  \bibfield  {author} {\bibinfo {author} {\bibfnamefont {J.}~\bibnamefont
  {Adam}} \emph {et~al.} (\bibinfo {collaboration} {ALICE}),\ }\href {\doibase
  10.1140/epjc/s10052-015-3422-9} {\bibfield  {journal} {\bibinfo  {journal}
  {Eur. Phys. J.}\ }\textbf {\bibinfo {volume} {C75}},\ \bibinfo {pages} {226}
  (\bibinfo {year} {2015})},\ \Eprint {http://arxiv.org/abs/1504.00024}
  {arXiv:1504.00024 [nucl-ex]} \BibitemShut {NoStop}%
\bibitem [{\citenamefont {Acharya}\ \emph {et~al.}(2019)\citenamefont {Acharya}
  \emph {et~al.}}]{Acharya:2018orn}%
  \BibitemOpen
  \bibfield  {author} {\bibinfo {author} {\bibfnamefont {S.}~\bibnamefont
  {Acharya}} \emph {et~al.} (\bibinfo {collaboration} {ALICE}),\ }\href
  {\doibase 10.1103/PhysRevC.99.024906} {\bibfield  {journal} {\bibinfo
  {journal} {Phys. Rev.}\ }\textbf {\bibinfo {volume} {C99}},\ \bibinfo {pages}
  {024906} (\bibinfo {year} {2019})},\ \Eprint
  {http://arxiv.org/abs/1807.11321} {arXiv:1807.11321 [nucl-ex]} \BibitemShut
  {NoStop}%
\bibitem [{\citenamefont {Sharma}\ \emph {et~al.}(2019)\citenamefont {Sharma},
  \citenamefont {Cleymans}, \citenamefont {Hippolyte},\ and\ \citenamefont
  {Paradza}}]{Sharma:2018jqf}%
  \BibitemOpen
  \bibfield  {author} {\bibinfo {author} {\bibfnamefont {N.}~\bibnamefont
  {Sharma}}, \bibinfo {author} {\bibfnamefont {J.}~\bibnamefont {Cleymans}},
  \bibinfo {author} {\bibfnamefont {B.}~\bibnamefont {Hippolyte}}, \ and\
  \bibinfo {author} {\bibfnamefont {M.}~\bibnamefont {Paradza}},\ }\href
  {\doibase 10.1103/PhysRevC.99.044914} {\bibfield  {journal} {\bibinfo
  {journal} {Phys. Rev.}\ }\textbf {\bibinfo {volume} {C99}},\ \bibinfo {pages}
  {044914} (\bibinfo {year} {2019})},\ \Eprint
  {http://arxiv.org/abs/1811.00399} {arXiv:1811.00399 [hep-ph]} \BibitemShut
  {NoStop}%
\bibitem [{\citenamefont {Vovchenko}\ \emph {et~al.}(2019)\citenamefont
  {Vovchenko}, \citenamefont {Dönigus},\ and\ \citenamefont
  {Stoecker}}]{Vovchenko:2019kes}%
  \BibitemOpen
  \bibfield  {author} {\bibinfo {author} {\bibfnamefont {V.}~\bibnamefont
  {Vovchenko}}, \bibinfo {author} {\bibfnamefont {B.}~\bibnamefont {Dönigus}},
  \ and\ \bibinfo {author} {\bibfnamefont {H.}~\bibnamefont {Stoecker}},\
  }\href {\doibase 10.1103/PhysRevC.100.054906} {\bibfield  {journal} {\bibinfo
   {journal} {Phys. Rev.}\ }\textbf {\bibinfo {volume} {C100}},\ \bibinfo
  {pages} {054906} (\bibinfo {year} {2019})},\ \Eprint
  {http://arxiv.org/abs/1906.03145} {arXiv:1906.03145 [hep-ph]} \BibitemShut
  {NoStop}%
\bibitem [{\citenamefont {Das}\ \emph {et~al.}(2017)\citenamefont {Das},
  \citenamefont {Mishra}, \citenamefont {Chatterjee},\ and\ \citenamefont
  {Mohanty}}]{Das:2016muc}%
  \BibitemOpen
  \bibfield  {author} {\bibinfo {author} {\bibfnamefont {S.}~\bibnamefont
  {Das}}, \bibinfo {author} {\bibfnamefont {D.}~\bibnamefont {Mishra}},
  \bibinfo {author} {\bibfnamefont {S.}~\bibnamefont {Chatterjee}}, \ and\
  \bibinfo {author} {\bibfnamefont {B.}~\bibnamefont {Mohanty}},\ }\href
  {\doibase 10.1103/PhysRevC.95.014912} {\bibfield  {journal} {\bibinfo
  {journal} {Phys. Rev.}\ }\textbf {\bibinfo {volume} {C95}},\ \bibinfo {pages}
  {014912} (\bibinfo {year} {2017})},\ \Eprint
  {http://arxiv.org/abs/1605.07748} {arXiv:1605.07748 [nucl-th]} \BibitemShut
  {NoStop}%
\bibitem [{\citenamefont {Wheaton}\ and\ \citenamefont
  {Cleymans}(2009)}]{Wheaton:2004qb}%
  \BibitemOpen
  \bibfield  {author} {\bibinfo {author} {\bibfnamefont {S.}~\bibnamefont
  {Wheaton}}\ and\ \bibinfo {author} {\bibfnamefont {J.}~\bibnamefont
  {Cleymans}},\ }\href {\doibase 10.1016/j.cpc.2008.08.001} {\bibfield
  {journal} {\bibinfo  {journal} {Comput. Phys. Commun.}\ }\textbf {\bibinfo
  {volume} {180}},\ \bibinfo {pages} {84} (\bibinfo {year} {2009})},\ \Eprint
  {http://arxiv.org/abs/hep-ph/0407174} {arXiv:hep-ph/0407174 [hep-ph]}
  \BibitemShut {NoStop}%
\bibitem [{\citenamefont {Kraus}\ \emph {et~al.}(2009)\citenamefont {Kraus},
  \citenamefont {Cleymans}, \citenamefont {Oeschler},\ and\ \citenamefont
  {Redlich}}]{Kraus:2008fh}%
  \BibitemOpen
  \bibfield  {author} {\bibinfo {author} {\bibfnamefont {I.}~\bibnamefont
  {Kraus}}, \bibinfo {author} {\bibfnamefont {J.}~\bibnamefont {Cleymans}},
  \bibinfo {author} {\bibfnamefont {H.}~\bibnamefont {Oeschler}}, \ and\
  \bibinfo {author} {\bibfnamefont {K.}~\bibnamefont {Redlich}},\ }\href
  {\doibase 10.1103/PhysRevC.79.014901} {\bibfield  {journal} {\bibinfo
  {journal} {Phys. Rev.}\ }\textbf {\bibinfo {volume} {C79}},\ \bibinfo {pages}
  {014901} (\bibinfo {year} {2009})},\ \Eprint {http://arxiv.org/abs/0808.0611}
  {arXiv:0808.0611 [hep-ph]} \BibitemShut {NoStop}%
\bibitem [{\citenamefont {Abelev}\ \emph
  {et~al.}(2014{\natexlab{b}})\citenamefont {Abelev} \emph
  {et~al.}}]{Abelev:2013haa}%
  \BibitemOpen
  \bibfield  {author} {\bibinfo {author} {\bibfnamefont {B.~B.}\ \bibnamefont
  {Abelev}} \emph {et~al.} (\bibinfo {collaboration} {ALICE}),\ }\href
  {\doibase 10.1016/j.physletb.2013.11.020} {\bibfield  {journal} {\bibinfo
  {journal} {Phys. Lett.}\ }\textbf {\bibinfo {volume} {B728}},\ \bibinfo
  {pages} {25} (\bibinfo {year} {2014}{\natexlab{b}})},\ \Eprint
  {http://arxiv.org/abs/1307.6796} {arXiv:1307.6796 [nucl-ex]} \BibitemShut
  {NoStop}%
\bibitem [{\citenamefont {Adam}\ \emph
  {et~al.}(2016{\natexlab{a}})\citenamefont {Adam} \emph
  {et~al.}}]{Adam:2015vsf}%
  \BibitemOpen
  \bibfield  {author} {\bibinfo {author} {\bibfnamefont {J.}~\bibnamefont
  {Adam}} \emph {et~al.} (\bibinfo {collaboration} {ALICE}),\ }\href {\doibase
  10.1016/j.physletb.2016.05.027} {\bibfield  {journal} {\bibinfo  {journal}
  {Phys. Lett.}\ }\textbf {\bibinfo {volume} {B758}},\ \bibinfo {pages} {389}
  (\bibinfo {year} {2016}{\natexlab{a}})},\ \Eprint
  {http://arxiv.org/abs/1512.07227} {arXiv:1512.07227 [nucl-ex]} \BibitemShut
  {NoStop}%
\bibitem [{\citenamefont {Adam}\ \emph
  {et~al.}(2016{\natexlab{b}})\citenamefont {Adam} \emph
  {et~al.}}]{Adam:2016bpr}%
  \BibitemOpen
  \bibfield  {author} {\bibinfo {author} {\bibfnamefont {J.}~\bibnamefont
  {Adam}} \emph {et~al.} (\bibinfo {collaboration} {ALICE}),\ }\href {\doibase
  10.1140/epjc/s10052-016-4088-7} {\bibfield  {journal} {\bibinfo  {journal}
  {Eur. Phys. J.}\ }\textbf {\bibinfo {volume} {C76}},\ \bibinfo {pages} {245}
  (\bibinfo {year} {2016}{\natexlab{b}})},\ \Eprint
  {http://arxiv.org/abs/1601.07868} {arXiv:1601.07868 [nucl-ex]} \BibitemShut
  {NoStop}%
\end{thebibliography}%

\end{document}